\documentclass[12pt,epsfig]{article}
\usepackage{amsmath}
\usepackage{graphicx}
\usepackage{epstopdf}
\usepackage{lscape}
\usepackage{longtable}
\usepackage{float}
\usepackage[usenames, dvipsnames]{color}
\usepackage{rotating}
\usepackage{tablefootnote}

\begin{document}
	\begin{center}
		\textbf {Quality check of a sample partition using multinomial distribution}\\
	\end{center}
	\begin{center}
		Dr. Soumita Modak$^{*}$\\
		Faculty,
		Department of Statistics\\
		University of Calcutta,
		Basanti Devi College\\
		147B, Rash Behari Ave, Kolkata- 700029, India\\
		Email: soumitamodak2013@gmail.com\\
		Orcid id: 0000-0002-4919-143X\\
		Homepage: https://sites.google.com/view/soumitamodak
	\end{center}
	Abstract: In this paper, we advocate a novel measure for the purpose of checking the quality of a cluster partition for a sample into several distinct classes, and thus, determine the unknown value for the true number of clusters prevailing the provided set of data. Our objective leads us to the development of an approach through applying the multinomial distribution to the distances of data members, clustered in a group, from their respective cluster representatives. This procedure is carried out independently for each of the clusters, and the concerned statistics are combined together to design our targeted measure. Individual clusters separately possess the category-wise probabilities which correspond to different positions of its members in the cluster with respect to a typical member, in the form of cluster-centroid, medoid or mode, referred to as the corresponding cluster representative. Our method is robust in the sense that it is distribution-free, since this is devised irrespective of the parent distribution of the underlying sample. It fulfills one of the rare coveted qualities, present in the existing cluster accuracy measures, of having the capability to investigate whether the assigned sample owns any inherent clusters other than a single group of all members or not. Our measure's simple concept, easy algorithm, fast runtime, good performance, and wide usefulness, demonstrated through extensive simulation and diverse case-studies, make it appealing. 
	
	keyword: Cluster Analysis; Checking for clusters; Quality assessment of a cluster-partition; Evaluation of clusters' number; Multinomial distribution. 
	
	\section{Introduction}
	Scientists driven by the quest for discoveries of unveiled truth are equipped with the unsupervised classification, where no prior information is known regarding a new sample of observations and the sources of variation in the data set are tried to be explored by finding meaningful clusters (Ruspini 1970; Hartigan 1975; McLachlan and Peel 2000; Everitt, Landau and Leese 2001; Kaufman and Rousseeuw 2005; Balcan, Liang and Gupta 2014; Modak 2021; Modak et al. 2022; Sureja, Chawda and Vasant 2022;  Modak 2023a, 2023d, 2023g). We study the situation where each member of data is clustered in exactly one group (Jain, Murty and Flynn 1999; Johnson and Wichern  2007; Modak et al. 2018; Modak 2019, 2023b, 2023f). Then, it becomes straightforward to analysis the physical origins behind the resulting clusters. However, neither it is easy to find out the unknown true value of the prevailing clusters (denoted by $K$) nor to verify whether a considered value of $K$ is correct in reality. This still remains a global issue in spite of various measures, based on diverse perspectives, having been proposed to conquer the challenge (Dunn 1974; Lewis and Gale 1994; Pakhira, Bandyopadhyay and Maulik 2004; Aljarah and Ludwig 2013; Silva, Melton and Wunsch 2020; Modak 2022, 2023c, 2023e). Moreover, the bulk of the highly acclaimed measures like Cali\'{n}ski
	and Harabasz index,  Dunn index, average silhouette width, nearest neighbor classification error rate or connectivity (Cali\'{n}ski \& Harabasz 1974, Dunn 1974, Rousseeuw 1987, Ripley 1996, Handl et al. 2005) are unable to investigate whether there is at all any heterogeneity in a sample or not, i.e., $K=1$ or $K>1$. Without checking so, all efforts to look for a value of $K$ (assumed to be $\geq2)$ would just lead us to spurious answer in case of a single-cluster scenario. On the other hand, very few owning this characteristic are either possessing heavy parametric assumptions and/or involving huge computations, e.g., the Bayesian Information Criterion, Bayes factor or gap statistic
	(Schwarz 1978; Frayley and Raftery 1998; Tibshirani, Walther and Hastie 2001). In this context, we advise a novel, simple, distribution-free and fast implemented index that checks the quality of a sample partition to estimate $K(\geq 1)$.
	
	Our measure owns the robustness with respect to the distributions of the original data or the cluster-populations. It is based on the multinomial distribution (McCullagh and Nelder 1989; Johnson, Kotz and Balakrishnan 1997) applied to the distances of all members, in a cluster, computed from their cluster representatives, wherein the category-wise probabilities represent the different positions of the members in a particular cluster with respect to its representative. Thus, it is free from the distributional assumptions regarding the sample population. In the present study, we consider the widely successful cluster indices Cali\'{n}ski
	and Harabasz index (Cali\'{n}ski \& Harabasz 1974), Dunn index (Dunn 1974), nearest neighbor classification error rate (Ripley 1996), gap statistic (Tibshirani, Walther and Hastie 2001), connectivity measure (Handl et al. 2005), and kernel-based index (Modak 2023e), to compare the performance of our devised metric. The proposed measure uses only the distances of members from their cluster representatives, instead of the massive interpoint distance matrix involved throughout in its competitors, whereas the gap statistic calls for a parametric distribution and concerns an extremely computationally-extensive algorithm. These make our index quiet faster compared to its contenders and hence is preferred for big data analyses. Our suggested method assesses the quality of a sample partition and thereby estimates the unknown number of clusters present in the underlying sample. Most importantly, our index possesses the rarest quality among the existing cluster assessment indices, shared by only the gap statistic from the rival measures under comparison, that is to check whether there is any clustering pattern prevailing a data set or not, which is extremely useful in practice when a new sample is looked into for possible grouping. Our designed criterion, falls under the category of internal cluster accuracy measures which use only the given data and its partition resulted in the performed clustering, is eligible for finding division in a new data set without any other information available on its existing grouping. The proposed method carries out a thorough exploration of individual clusters independently in our advised way, and then the corresponding statistics are combined into a final measure capable of the targeted purpose, even in cases of the challenging separation of entangled, overlapping or closely-spaced clusters. Its easy lay-out, simple interpretation and superior performance, studied in terms of extensive simulation study and real-world data analyses, are discussed as follows. 
	
	Section 2 describes the construction of our measure, its properties, computation, significance and interpretation. Section 3 demonstrates the measure's application through different interesting case studies and its comparison with the competitors, with Section 4 concluding.
	\section{Method}
	Suppose we are given $N$ members from a sample into $K(\geq 1)$ hard clusters $C_1,...,C_K$ of respective sizes $N_1,...,N_K$ $\ni\sum_{k=1}^{K}N_k=N$. First, we determine a cluster representative (denoted by $R_k$ for the $k$-th cluster for $k=1(1)K$; namely, centroid or medoid) for individual clusters. Then, for every cluster, we compute the distances of all members, clustered in it, from their representative, and normalize them to lie in $[0,1]$ wherein the lower bound is attained when the cluster-representative coincides a member from the cluster itself. Next, the interval $[0,1]$ is divided into $l(>1)$ number of equally spaced sub-intervals, which are mutually exclusive, exhaustive and increasing in values, look like: 
	\begin{equation*}
		\biggl[\frac{0}{l}=0,\frac{1}{l}\biggl),\biggl[\frac{1}{l},\frac{2}{l}\biggl),...,\biggl[\frac{l-2}{l},\frac{l-1}{l}\biggl),\biggl[\frac{l-1}{l},\frac{l}{l}=1\biggl],
	\end{equation*}
	where we count the frequencies of the distances falling in the individual sub-intervals.
	
	Mathematically speaking, for $M_{mk}$ being the $m$-th member from the $k$-th cluster $C_k$, we compute
	\begin{equation}
		S_k=\{d(M_{mk},R_k), m=1,...,N_k\},
	\end{equation}
	where $d$ representing a distance measure, specified by the user. In an attempt of computational ease and avoidance of spareness, we consider the following set of distances:
	\begin{equation}
		S^{Nor}_k=\biggl\{\frac{d(M_{mk},R_k)}{\underset{m=1(1)N_k}{\max} d(M_{mk},R_k)}, m=1,...,N_k\biggl\},
	\end{equation} 
	through normalization $\ni$ all distances from $S^{Nor}_k\subseteq [0,1]$ (0 is attained for $R_k\in C_k$), and then it is divided into $l$ sub-intervals as stated above.
	\subsection{Implementation of multinomial distribution}	
	Let $N_{k}=$ the number of members $\in C_k$ and $N_{j|k}=$ the number of members $\in C_k$ falling in the $j$-th sub-interval from $S_k^{Nor}$. Obviously, $0\leq N_{j|k}\leq N_k$ for any $l$ chosen. Now, 
	\begin{equation}\label{bin}
		N_{j|k}\sim \text{bin}(N_k,\pi_{j|k}),
	\end{equation} 
	where ``bin" stands for a binomial distribution and $\pi_{j|k}=$ probability for a member from $C_k$ falling in the $j$-th sub-interval, for a fixed $k\in\{1,...,K\}$ and  $j=1,...,l$. Then, for every $k$, 
	\begin{equation}\label{MN}
		\mathbf{N}_k=(N_{1|k},...,N_{l|k})'\sim\hspace{.05in} \text{MN $(N_k,l,\pi_{1|k},...,\pi_{l|k})$},
	\end{equation}
	where ``MN" stands for a multinomial distribution (see, McCullagh and Nelder 1989; Johnson, Kotz and Balakrishnan 1997), here satisfying\\
	(i) 
	\begin{equation}
		E(N_{j|k})=N_k\pi_{j|k}
	\end{equation} 
	with $\sum_{j=1}^{l}\pi_{j|k}=1$ and $\sum_{j=1}^{l}N_{j|k}=N_k$, and\\
	(ii) \begin{equation}\label{dispersion}
		Disp(\mathbf{N}_k)=N_k\{diag(\boldsymbol{\pi}_k)-\boldsymbol{\pi}_k\boldsymbol{\pi}_k'\}=\Sigma_k\hspace*{.05in}\text{(say)},
	\end{equation} 
	where $\boldsymbol{\pi}_k=(\pi_{1|k},...,\pi_{l|k})'$ constitutes the diagonal entries for the diagonal matrix: $diag(\boldsymbol{\pi}_k)$, i.e. the above equation can be explicitly written as:
	\begin{equation*}
		\Sigma_k^{l\times l}=\begin{bmatrix}
			N_k\pi_{1|k}(1-\pi_{1|k})&-N_k\pi_{1|k}\pi_{2|k}&...&-N_k\pi_{1|k}\pi_{\overline{l-1}|k}&-N_k\pi_{1|k}\pi_{l|k}\\
			-N_k\pi_{2|k}\pi_{1|k}&N_k\pi_{2|k}(1-\pi_{2|k})&...&-N_k\pi_{2|k}\pi_{\overline{l-1}|k}&-N_k\pi_{2|k}\pi_{l|k}\\
			\vdots&\vdots&...&\vdots&\vdots\\
			
			-N_k\pi_{l|k}\pi_{1|k}&-N_k\pi_{l|k}\pi_{2|k}&....&-N_k\pi_{l|k}\pi_{\overline{l-1}|k}&N_k\pi_{l|k}(1-\pi_{l|k})\\
		\end{bmatrix}.
	\end{equation*}
	\subsection{Our rationale}	
	A cluster representative is the most typical observation (vector) within the cluster, which can be either an original member from given sample, like the medoid in $K$-medoids clustering (Kaufman and Rousseeuw 2005; Modak et al. 2020), or some function computed over each member of the cluster, e.g., centroid in $K$-means algorithms (MacQueen 1967; Hartigan 1975), revealing the inherent properties of the cluster best. Therefore, we study how far the members of a cluster are located from an appropriately determined cluster-representative. In checking the quality of a partition for a designated sample, we analyze the dispersion of the clusters with respect to their respective representatives, which is quantified in terms of the distances computed between individual members of a cluster and its representative.  
	The more members of a cluster cloud around its typical member  (i.e., cluster-representative) and the less members lying far from it, the better is the quality for the clustering; in which situation we expect more information by means of cluster members or equivalently more dispersion in the cluster near its representative, compared to the areas increasing in distance from the cluster-representative. 
	
	To check the above, we divide the interval [0,1] for the normalized distances into $l$ (user-defined) mutually exclusive and exhaustive sub-intervals of equal length. For cluster $k$, the frequency in the $j$-th sub-interval is a random variable as per Eq.~\eqref{bin}, where the joint distribution of the frequencies from all sub-intervals is assumed to be a multinomial, as explained by Eq.~\eqref{MN} in Section 2.1. Clustering is an exploratory data analysis, where we neither know the values for these random variables in advance while clustering a sample into a fixed number of partitions, nor the implementation of an arbitrary clustering algorithm guarantees constant values for them even with a specified $K$. Therefore, our approach is justified, and quite robust in the sense that in spite of adopting a parametric procedure in terms of multinomial distribution, the analysis is distribution-free due to the fact that our method does not depend on the population-distribution of the sample.
	
	The above-mentioned quality check of cluster $k$, from an available partition for the sample, is now interpreted in terms of each individual $N_{j|k}$, with the $k$ specified, as follows:\\
	(a) If
	\begin{equation}
		Var(N_{1|k})>...>Var(N_{l|k}) \hspace*{.05in} \text{holds},
	\end{equation}
	then the quality of the cluster $k$ is a best possible, whereas\\
	(b) when 
	\begin{equation}
		Var(N_{1|k})<...<Var(N_{l|k}) \hspace*{.05in} \text{happens},
	\end{equation}
	the opposite occurs indicating the worst quality of the cluster.\\
	This same explanation is valid for every cluster $k=1(1)K$, and based on that we proceed to propose our novel measure in the following section. 
	\subsection{Proposed measure}
	A plausible measure for quality check of the $k$-th cluster:
	\begin{equation}\label{stat}
		C_{MN_{k}} =\sum_{j=1}^{l}w_{j|k}Var(N_{j|k}),
	\end{equation}
	where $w_{j|k}$ s (``s" letter is added to mean plural) are non-negative weights selected for giving relative importance to $Var(N_{j|k})$ s $\ni w_{j|k}\downarrow$ in $j$.
	
	A simple choice of $w$ s fulfilling the aforesaid requirements, in the formation of our quality check index, would be: 
	\begin{equation}
		w_{j|k}=	l+1-j,\forall j=1(1)l, 
	\end{equation} 
	maintained throughout all clusters $k=1(1)K$. We report the results for our case studies in Section~3 with this selection.
	
	A large value for our proposed criterion (Eq.~\ref{stat}) is desired, whose analytical expression using the distributions of $N_{j|k}$ s (Eq.~\ref{dispersion}) is obtained as:
	\begin{equation}
		C_{MN_k}=\sum_{j=1}^{l}w_{j|k}N_k\pi_{j|k}(1-\pi_{j|k}),
	\end{equation}
	where the unknown parameters $\pi_{j|k}$ s are replaced with their respective unbiased estimators
	\begin{equation*}
		\hat\pi_{j|k}= N_{j|k}/N_k
	\end{equation*}
	with $\sum_{j=1}^{l}\hat{\pi}_{j|k}=1$, and we reach our (estimated) measure (i.e., sample version) for cluster $k$ as:
	\begin{equation}\label{measure}
		\hat{C}_{MN_k}=\sum_{j=1}^{l}w_{j|k}N_k\hat{\pi}_{j|k}(1-\hat{\pi}_{j|k}).
	\end{equation}
	
	Now, we propose our final measure, for checking the quality of the entire sample partition, as the sum of the statistics $\hat{C}_{MN_k}$ s computed for each cluster $k$ (Eq.~\ref{measure}):
	\begin{align}\label{finalexp}
		C^K_{MN}&=\nonumber\sum_{k=1}^{K}\hat{C}_{MN_k} \\ \nonumber
		&=\sum_{k=1}^{K}\sum_{j=1}^{l}w_{j|k}N_k\hat{\pi}_{j|k}(1-\hat{\pi}_{j|k}) \\ \nonumber
		&=\sum_{k=1}^{K}\sum_{j=1}^{l}w_{j|k}\frac{N_{j|k}}{N_k}\Biggl(\sum_{j'(\neq j)=1}^{l}N_{j'|k}\Biggl),\\
	\end{align}
	wherein if for a $K$, the data own one or more singleton cluster(s), then we suggest contributing zero to Eq.~\eqref{finalexp} for each such cluster.
	
	\subsubsection{Quality check and determination of the number of clusters}
	It is easy to check, theoretically, that our proposed measure, empirically computed based on the designated sample-partition, is an unbiased estimator of its population version (approximately), because:
	\begin{align}
		E(C^K_{MN})&=\nonumber\sum_{k=1}^{K}\{E(\hat{C}_{MN_k})\}\\ \nonumber
		&=\nonumber\sum_{k=1}^{K}\Biggl[\sum_{j=1}^{l}w_{j|k}E\{\widehat{Var}(N_{j|k})\}\Biggl]\\ \nonumber
		&=\nonumber\sum_{k=1}^{K}\sum_{j=1}^{l}w_{j|k}N_kE\big\{\hat{\pi}_{j|k}(1-\hat{\pi}_{j|k})\big\}\\ \nonumber
		&=\nonumber\sum_{k=1}^{K}\sum_{j=1}^{l}w_{j|k}(N_k-1)\pi_{j|k}(1-\pi_{j|k})\\ \nonumber
		&\simeq\sum_{k=1}^{K}\sum_{j=1}^{l}w_{j|k}N_k\pi_{j|k}(1-\pi_{j|k})\\ \nonumber
		&= C^K_{MN},\\
	\end{align}
	which is a coveted characteristic of our new measure guaranteeing the reliability of its value, computed based on the given partition of a sample, to check the quality of the same.
	
	The index suggested $C^K_{MN}\geq 0$ $\ni$
	\begin{equation*}
		C^K_{MN}\uparrow\hspace*{.05in} <=>\hspace*{.025in}\text{partition quality}\hspace*{.025in} \uparrow,
	\end{equation*}
	wherein $K$ is estimated over a range of  considered values $\{1,2,...,K_{\max}\}$ as
	\begin{equation}\label{Khat}
		\hat{K}=\bigg\{K:\underset{1\leq K\leq K_{\max}}{\max}\hspace*{.05in} \big\{C^K_{MN}\big\}=C^{\hat{K}}_{MN}\bigg\}.
	\end{equation}
	$C^K_{MN}$ attains the lower boundary value zero only if there exist one sub-interval containing all observations and the others having zero for all clusters, which is unlikely to happen. The zero value can occur numerically, as per our approach, when all the resultant clusters are singletons for $K=N$; however, in practice we are concerned with a value of  $K<<N$.
	\subsubsection{Hyperparameters}
	Our measure has two tuning parameters or hyperparameters as $K$ and $l$. In association with the clustering algorithms (like $K$-means, $K$-medoids; see, Hartigan 1975; Kaufman and Rousseeuw 2005) for which $K$ is provided as a priori, we vary $K=2,3,...,K_{\max}$ ($K_{\max}$ depends on the case-study; we select six for all cases in Section 3); whereas the other kind of algorithms evaluates $K$ during the cluster analysis itself (e.g., DBSCAN, kernel-based clustering methods; see, Ester et al. 1996; Matioli et al. 2018; Modak 2023a). Whatever be the procedure, the final value of $K$ is estimated by the quality-check property of our suggested measure, which hints at a best possible clustering with its highest value. 
	
	Moreover, if we need to examine the sample for the presence of more than one clusters, we can compare the values of $\{C^K_{MN},K=2(1)K_{\max}\}$ with the value of $C^1_{MN}$ (see, Eq.~\ref{finalexp}), in case of the first kind of clustering algorithms; and for the other kind, we make the comparison between $C^{K'}_{MN}$, wherein $K=K'$ is determined by the algorithm, and $C^1_{MN}$. Our index for a single-group $C^1_{MN}$ is calculated over all members of the sample, with an analogously defined sample-representative for the entire data set as the cluster-representatives are constructed for each individual clusters with $K$ greater than one in the implemented algorithm. Finally, in the former case, the verdict is made by Eq.~\eqref{Khat} as depicted in Section 2.3.1; whereas the latter case selects $\hat{K}$ as either one or $K'$ whichever produces $\max\{C^1_{MN},C^{{K'}}_{MN}\}$.
	
	Another hyperparameter $l$ is a non-negative integer-valued constant being at least 2 (usually much greater than 2 to expose the underlying structure well) involved in our cluster assessment measure. The numerical value of the suggested index depends on $l$, whose value is case-dependent and provided by the user, should be varied. We go for the usual trial-and-error method, generally adopted in such exploratory studies with no global objective approach in existence (Silverman 1986; Bandyopadhyay and Modak 2018; Matioli et al. 2018; Modak 2022). $l$ is expected to produce robust answers with natural resultant clusters for all plausible values of it (see, case study (1) in Section 3). On the other hand, for a challenging clustering pattern, different values of $l$ may lead to distinct answers, where a best value generates the highest value for our measure corresponding to an optimal partition in accordance with the index's intrinsic design (e.g., see, case study (6) from Section 3).
	
	\section{Case studies}
	The efficacy of our measure is to be evaluated in checking the quality of an available partition for the assigned sample and thereby in estimating the true unknown number of clusters present in the data set. To carry out so, the sample is first clustered using an appropriate cluster algorithm and we then compute our cluster accuracy index along with its well-established rivals Cali\'{n}ski and Harabasz index (CH; Cali\'{n}ski and Harabasz 1974), Dunn index (Dunn; see, Dunn 1975), $\tilde{K}$-nearest neighbor classification error rate ($\tilde{K}$NN; Ripley 1996), gap statistic (Gap; Tibshirani, Walther and Hastie 2001), connectivity (Conn; Handl et al. 2005), and kernel-based index ($M_{clus}$; Modak 2023e). Dunn is defined as a ratio of the smallest distance between
	members from separate clusters to the largest within-cluster distance, and CH is designed as a ratio of between-cluster mean of distances to within-cluster mean of distances, where both the indices are greater than zero with increasing values indicating higher-quality partition for a sample. A sophisticated measure gap statistic, constructed on the concept of the within-cluster-dispersion, is a real-valued index producing the optimum with its maximum value. Recently invented kernel density-based cluster validity index, with bandwidth $h$ (a hyperparameter, here $h=h^{*}$ for the classical Gaussian kernel, for details, see Modak 2023e) for the involving kernel, can be considered as a robust developed version of the popular cluster quality measure average silhouette width (Rousseeuw 1987). It accepts any value from -1 to 1 with a larger one hinting at a superior classification.  On the other hand, Conn measuring the tightness for the clusters, is a kind of sum of proximity between each member and its $J$ (a tuning parameter, here $J=10$, unless mentioned otherwise, is chosen as a trade-off between required information and computation burden) nearest neighbor(s). Another classical nearest neighbor-based measure is $\tilde{K}$NN that accounts for the agreement between cluster memberships of a member and its $\tilde{K}$ (a tuning parameter, here $\tilde{K}=9$, except stated otherwise, taken is an odd number in order to avoid the randomness in the decision to be made in case a tie is occurred in terms of the considered nearest neighbors) nearest neighbor(s). Both of the last statistics accept non-negative values where smaller means higher-quality clustering. 
	
	All indices are a function of $K$ and depend on a distance measure like ours, where we consider the commonly used Euclidean metric unless cited otherwise; nevertheless, none of the rivals, except Gap, can check a sample for the presence of clustering (i.e., whether $K=1$ or greater) but ours. It is to be noted that, throughout the case studies, we generally consider the cluster representatives to be the simple mean vectors, with entries as the dimension-wise arithmetic means computed over all members in an individual cluster; unless any special case happens when they are mentioned to be different according to the compatibility with the corresponding situation. 
	
	Comparison of the time complexity among the measures, with specified hyperparameter in the concerning measure, underlines: for the $k$-th cluster consisting of $N_k$ members, our measure involves $N_k$ distances or $(N_k-1)$ distances if the cluster-representative itself is a member of the cluster, on the other hand, the other measures need ${N_k\choose 2}$ interpoint distances; in addition, the gap statistic adopts the technique of bootstrapping, that needs a large number of bootstrap samples, which is quite time-consuming even with the help of well-known in-built function developed by computer languages.
	
	For numerical computation, we use the statistical programming language `R' version 4.2.2 using a 64-bit laptop with core i3 processor. The corresponding code is given in the appendix for the interested readers, where we explain the computation of our measure with $K=2(1)K_{\max}(=6)$ for a part of the following case study. If $C^1_{MN}$ is required, we simply find Eq.~\eqref{finalexp} with $K=1$, i.e. no clustering is performed and the whole set of given observations is considered as a single cluster, where the cluster-representative is computed over the entire data set. In order to provide the exact runtime of our proposed measure, we study so at the end of the first case study under different values of dimension, size, number of clusters present in the data set considered for cluster analysis, and the hyperparameter $l$ involved in our index.
	
	Case study (1): We consider a multivariate normal distribution in a 10-dimensional space with the components connected through a $t$-copula (Nelsen 2006; Modak and Bandyopadhyay 2019), which is explained by a 10-variate $t$-distribution with 2 degrees of freedom and the following ${10 \times 10}$ correlation matrix: 
	\begin{equation*}
		\begin{bmatrix}
			1&0.15&...&0.15&0.15\\
			0.15&1&...&0.15&0.15\\
			\vdots&\vdots&...&\vdots&\vdots\\
			0.15&0.15&...&1&0.15\\
			0.15&0.15&....&0.15&1\\
		\end{bmatrix}.
	\end{equation*}
	The copula can be analytically expressed as:
	\begin{equation}
		C(\mathbf{u})=F[F^{-1}_1(u_1),...,F^{-1}_{10}(u_{10})], 
	\end{equation}
	where 
	\begin{equation*}
		\mathbf{u}=(u_1,...,u_{10})', 0<u_i<1\forall i=1(1)10,
	\end{equation*}
	and $F$ is the joint cumulative distribution function (cdf) for the multivariate $t$-distribution with $F^{-1}_i$ as the inverse function of the cdf for the marginal distribution of the $i$-th component. Our synthetic data set under the above compound structure is created with three inherent groups of unequal sizes 45, 50, and 70, with the mean vectors $0\times \mathbf{1}_{10}, -3\times \mathbf{1}_{10}$, and $3\times \mathbf{1}_{10}$, respectively ($\mathbf{1}_{10}$ symbolizes a 10-variate vector with all entries equal to 1).
	
	In this situation, we apply the widely used partitioning-based clustering method  $K$-means (``Hartigan-Wong" algorithm; see, Hartigan 1975, Modak et al. 2018), as it is well-known for exploring the Gaussian clusters. In order to implement our cluster accuracy measure, we select the cluster representatives to be the cluster-wise means, since the clustering algorithm aims at minimizing the within-cluster sum of squares with respect to the means of the clusters. The computed values of our index (for $l=10$ chosen) along with its competitors, for $K$-means clustering, are given in Table~\ref{t1:MulNor}, where the well-separated clusters are prominently exposed by all the indices.
	
	To check the robustness of the clustering results, we perform another classical and more robust clustering method $K$-medoids (``PAM" algorithm; see, Kaufman and Rousseeuw 2005; Modak et al. 2017, 2020), as it is specially designed for spherically shaped groups. Computation of our measure uses medoids for individual groups as the cluster representatives in accordance with the present clustering algorithm. Table~\ref{t1:MulNor} shows all measures (except $M_{clus}$) direct us to $\hat{K}=3$ with desired success. 
	
	It proves that our measure is capable enough to disclose the true clustering pattern inherent to a sample, whereas its different values for the two clustering algorithms indicate distinct outcome by them. Therefore, to compare their performances, we utilize the original class labels available for all members from this synthetic data set (see, Table~\ref{t2:MulNor}). From the table it is clear that a higher value of our index is an evidence for better clustering (here $K$-means), as per its intrinsic design. It justifies the construction and interpretation of our approach empirically. 
	
\begin{figure}
		\centering
		\includegraphics[width=1\textwidth]{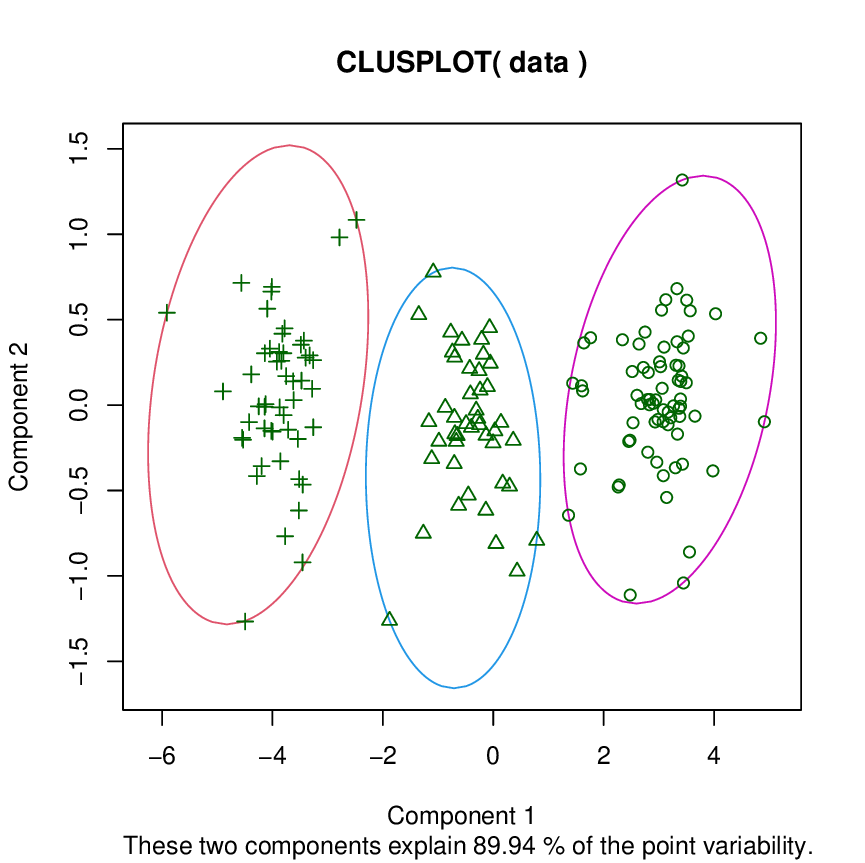}
		\caption{Three clusters, resulted in $K$-means clustering for a multivariate normal data set from case-study (1), in terms of the first two linear and orthogonal principal components with 89.94\% variation, are shown (different clusters plotted with distinct patterns), wherein clusters are outlined by different-colored  spheres.}\label{f_MulNor}
	\end{figure}

	Now, we consider the impact of the hyperparameter $l$ on our measure and hence on the output produced by it. As $K$-means algorithm is better suited for the present data (from Table~\ref{t2:MulNor}), we study the values of our measure with varying $l$, in association with $K$-means for different $K$ in Table~\ref{t3:MulNor}. We conclude the consistency of our measure, with respect to the various values of its tuning parameter $l$, in exploring the existent natural clusters (see, Fig.~\ref{f_MulNor}).
	
	Next, we consider the most important aspect regarding the clustering of a sample, which is to investigate whether there is any further clustering pattern existing in the data at all other than the set of data as a whole. Generally, the cluster indices, through which $K$ is determined in association with a clustering algorithm, are not capable of this task. As a result, if we assume $K\geq 2$ and start looking for $\hat{K}$, we would never be able to explore the original scenario while true value of $K$ is one. This serious issue is solved by our method, where we compute the value of the Eq.~\eqref{finalexp} for $K$ specified as one. The present data possess: $C^{1}_{MN}=788.5697$ (for $l=10$ along with $K$-means), when compared with our index's counterpart for $K$ greater than one: $\big\{C^K_{MN},K=2(1)6\big\}$ (see, Table~\ref{t1:MulNor}), prominently establishes the original three clusters.
	
	As we have generated the clusters from known probability distributions, we are in a position to provide the expected probability of correct recovery by the proposed measure through a Monte Carlo simulation technique, where we consider 1000 Monte Carlo replications (the largest ever used for cluster analysis, to our knowledge). 
	For our index, we count the success times (say, $I)$ over different $l$ as: $I=(834, 921, 944, 944, 939, 950, 946, 944, 939)$ for $l=(5, 7,10,11,12, 13, 14, 15, 17)$, simply by a grid-search method as depicted in the form of the following matrix. Here a row with $l$ values has the succeeding row for the corresponding $I$ values, where the ones (marked in bold) are computed in a particular stage, out of a total of four $S1-S4$, used to reach our decision in this case:  
	$$\begin{pmatrix}
		&&&&&S1&&&&\\
		l& 5 & 7& \textbf{10} &11& 12& 13& 14& 15& 17\\
		I&834& 921& \textbf{944}& 944& 939& 950& 946& 944& 939\\
		&&&&&S2&&&&\\
		l&\textbf{5} & \textbf{7}& 10 &11& 12& \textbf{13}& 14& 15& \textbf{17}\\
		I&\textbf{834}& \textbf{921}& 944& 944& 939& \textbf{950}& 946& 944& \textbf{939}\\
		&&&&&S3&&&&\\
		l&5 & 7& 10 &\textbf{11}& 12& 13& 14& \textbf{15}& 17\\
		I&834& 921& 944& \textbf{944}& 939& 950& 946& \textbf{944}& 939\\
		&&&&&S4&&&&\\
		l& 5 & 7& 10 &11& \textbf{12}& 13& \textbf{14}& 15& 17\\
		I&834& 921& 944& 944& \textbf{939}& 950& \textbf{946}& 944& 939\\
	\end{pmatrix}.$$
	This shows the consistency of the answers with respect to $l$; whereas for $l<10$, the loss of information starts affecting the same. We report $I=950$ for $l=13$, with corresponding results of its rival measures in Table~\ref{t:MCS}. Our index becomes either superior or competitive in comparison with its challengers.
	
	As far as the computation time of our metric is concerned, we consider multivariate normal data sets under the above-specified $t$-copula set-up from the following combinations of different values for: (a) dimension of data, highest at 30, (b) number of clusters, up to $K=20$, where the adjacent clusters are formed at a difference of 1 in all entries of their mean vectors, (c) equal size of each cluster (i.e. $N_1=...=N_K$), maximum of size 500 (i.e. largest of data set has the size $N=500\times 20= 10,000)$, and (d) hyperparameter $l$ till 30. For clustering, we implement $K$-medoids algorithm, for which the R code, provided in the appendix, can be readily used with changes under (a)-(d). Using this partition, into $K$ (designated under (b) ) groups, when we compute our measure $C^K_{MN}$, it takes just less than a second, which shows fast runtime of the suggested index and its preference for big data analyses.
	
	Case study (2): The second study is made to confirm the usefulness of our measure in case of a single cluster. For this objective, we simulate a multivariate sample under the same structure as in case study (1), with only difference in the number of clusters. Here we study one cluster, instead of three, where all observations are possessing the mean vector $0\times \mathbf{1}_{10}$. In association with $K$-means clustering,
	our index (for $l=10$) successfully discovers the single cluster (see, Table~\ref{t:noclustering}).
	
	Estimation of the probability for the above by Monte Carlo simulation leads to: $(I,l)=(892,5),(907,7),(961,10),(968,13),( 968,15), (960,17)$, wherein we see similar kind of impact of $l$ on $I$ as in the last case study. It manifests our index's effectiveness to identify single-cluster situation with a great degree and consistency over different plausible values of $l(\geq 10)$; whereas no other competitor, except gap statistic, shares this vital property of single-group exploration, that appears to be competitive to our measure (see, Table~\ref{t:MCS}).
	
	Case study (3): Gaussian-cluster analyses are followed by non-Gaussian clusters in this study, where 3 groups (at difference of three between the dimension-wise means of nearest group duo, as in case study 1) with equal size of 50 are drawn from a four-dimensional non-normal population (Vale \& Maurelli 1983). The multivariate structure is built by the correlation matrix with all off-diagonal elements 0.6. Deviation from normality is ascertained by providing dimension-wise skewness ($b_1$) = 1.75 and kurtosis ($b_2$) = 2 (Joanes \& Gill 1998). 
	
	We carry out the cluster analysis using the agglomerative hierarchical algorithm, with the average linkage which is thought to be producing reasonably robust answer for non-spherical clusters (Kaufman and Rousseeuw 2005, Modak 2023d). Results over 1,000 replications are given in Table~\ref{t:MCS}. As usual, initially we compute our measure for $l=10$ (giving rise to $I=535$), and then, as gathered from the earlier studies, avoiding the loss of information, we check for $l\in\{15,20,25,30,35\}$ leading to corresponding results:
	$I=(633 , 676 , 686 , 703 ,698)$. For further possible improvement, we look into $l=29$ and $31$ giving respective $I$ values as 694 and 697. It shows an optimal answer is obtained for $l=30$ which is reported in the table along with its competitors. This case study needs larger values of $l$, compared to the previous ones, to reach consistently good outcome through our index. It suggests the complexity of the underlying clusters; however, our measure stands significantly better than most of its contenders.
	
	Case study (4): Next study is carried out under the non-normal set-up of the last and through the same clustering algorithm, but concerning a single cluster. Our advised index with $l\in\{10,13,15,20,25,30,35\}$ gives rise to respective $I=(775, 880, 939 , 961, 970, 974,968)$, which significantly outperforms the gap statistic (referred to Table~\ref{t:MCS}).
	
\begin{figure}
		\centering
		\includegraphics[width=1\textwidth]{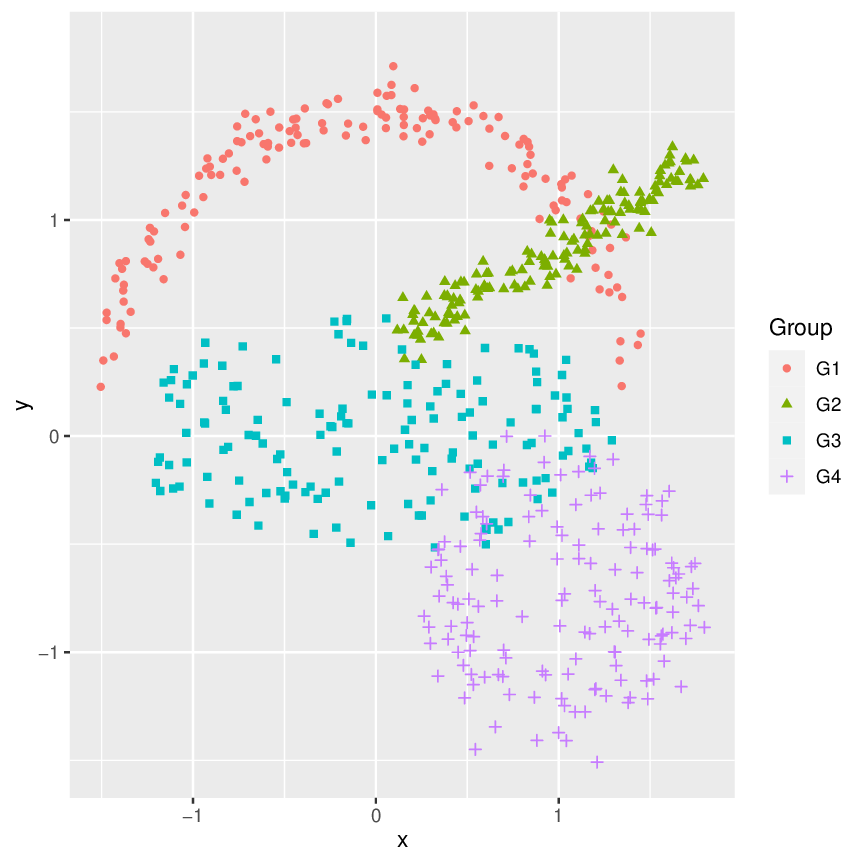}
		\caption{Arbitrary-shaped synthetic groups G1-G4 with noisy observations from case study (5).}\label{f1_ArShCl}
	\end{figure}

	Case study (5): After demonstration with respect to the identically shaped clusters, now we generate a more realistic and challenging scenario with four groups, namely G1-G4 each of size 150, having arbitrary shapes with noisy (by additive Gaussian noise with mean zero and standard deviation 0.075) and overlapping observations (see, Fig.~\ref{f1_ArShCl}, wherein for the purpose of an effortless visual depiction, we consider bivariate observations).
	First group G1 is consisted of data on the arc of a semi-circle and G2 is created using the equation of a line within a specified region. Next, G3 is utilizing bivariate points from the entire area of an ellipse, whereas G4 is of a circle.
	
	Here we carry out the popular density-based clustering method DBSCAN using a kd-tree (Ester et al. 1996; Campello et al. 2013), established for uncovering such differently shaped clusters contaminated with noises.
	Unlike $K$-means, $K$-medoids or hierarchical algorithms, it does not need $K$ to be specified. Because $K$ is estimated by the algorithm itself. It has two tuning parameters $\epsilon$ and $Minpts$ whose values are very crucial to obtain a plausible clustering outcome. Selection of their values is neither easy nor objective; however, a commonly adopted thumb-rule is to choose $Minpts=2\times p$ (where $p=$ dimension of data), for which $\epsilon$ is that value where a knee occurs in the curve of the
	$Minspts$-nearest neighbor distance plot (Hahsler, Piekenbrock,
	and Doran 2019). Nevertheless, in practice, the values should be varied before the final decision. 
	
\begin{figure}
		\centering
		\includegraphics[width=1\textwidth]{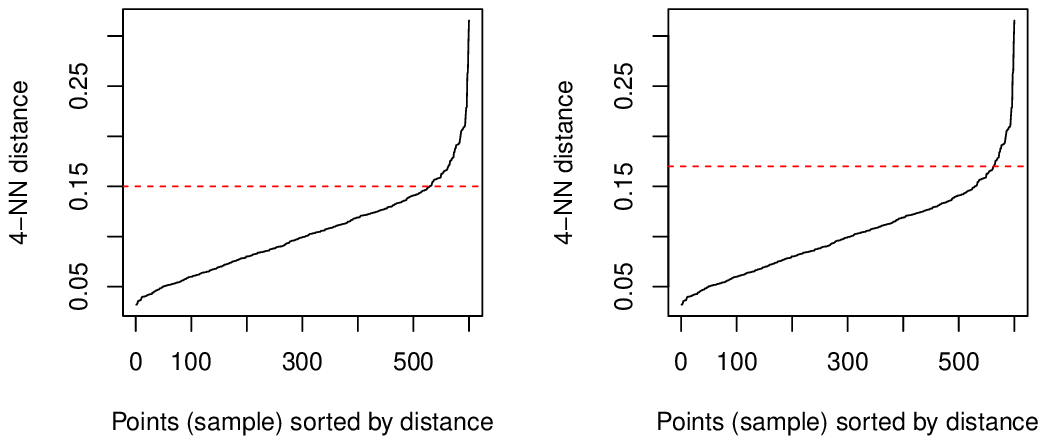}
		\caption{4-nearest neighbor (NN) distance plot of the sample from case study (5), considered for selection of the tuning parameters of DBSCAN, for $Minpts=4$ along with two different values of $\epsilon=0.15$ (left) and $0.17$ (right).}\label{f2N3_ArShCl}
	\end{figure}

	For example, different values of the hyperparameter duo, i.e. $(Minspts,\epsilon)=(4,0.15)$ and $(4,0.17)$ (see, Fig.~\ref{f2N3_ArShCl}), evaluate $\hat{K}=5$ and $4$. The latter gives off better partition of the sample, which is manifested in Table~\ref{t:ArShCl} by all the indices, except Gap which is not compatible with such comparison. It proves the success of our proposed measure in indicating the true number of arbitrary-shaped clusters with noisy observations. This does it robustly for both types of cluster-representatives as: (i) the cluster-means (commonly used dimension-wise mean; see, third col of Table~\ref{t:ArShCl}), and (ii) the cluster-modes (i.e., dimension-wise univariate mode, given in fourth col of Table~\ref{t:ArShCl}, as the implemented algorithm is density-based).
	
	Case study (6): After simulated samples, it is time to illustrate the effectiveness of our measure in clustering real-life data. For this, we choose the ``mtcars" \footnote{retrieved from `mtcars' data set embedded in `R' programming language} data set (Henderson and Velleman 1981), where the data are collected from the 1974 Motor Trend US magazine for 32 different car models on the following four components: (i) am:	transmission (automatic/manual), (ii) cyl: number of cylinders, (iii) hp:	gross horsepower, and (iv) wt:	weight (per 1000 lbs). It makes quite an interesting data set having measurements on mixed scales, where the first component is actually an attribute having two categories with the rest as numerical variables.
	
	First, the two categories of ``am" automatic and manual are converted to zero and one, respectively, and we obtain am in the form of a binary variable. Then, we fit the logistic regression model for it on the other variables, namely, cyl, hp, and wt (McCullagh and Nelder 1989; Dobson 1990; Agresti 2002). The Wald's test reveals that the only variable affecting am significantly is wt, which leads to the cluster analysis in the bivariate space of am-wt using $K$-medoids algorithm which adopts the Gower's distance for these mixed-scaled data (Gower 1971; Kaufman and Rousseeuw 2005).  
	
\begin{figure}
		\centering
		\includegraphics[width=1\textwidth]{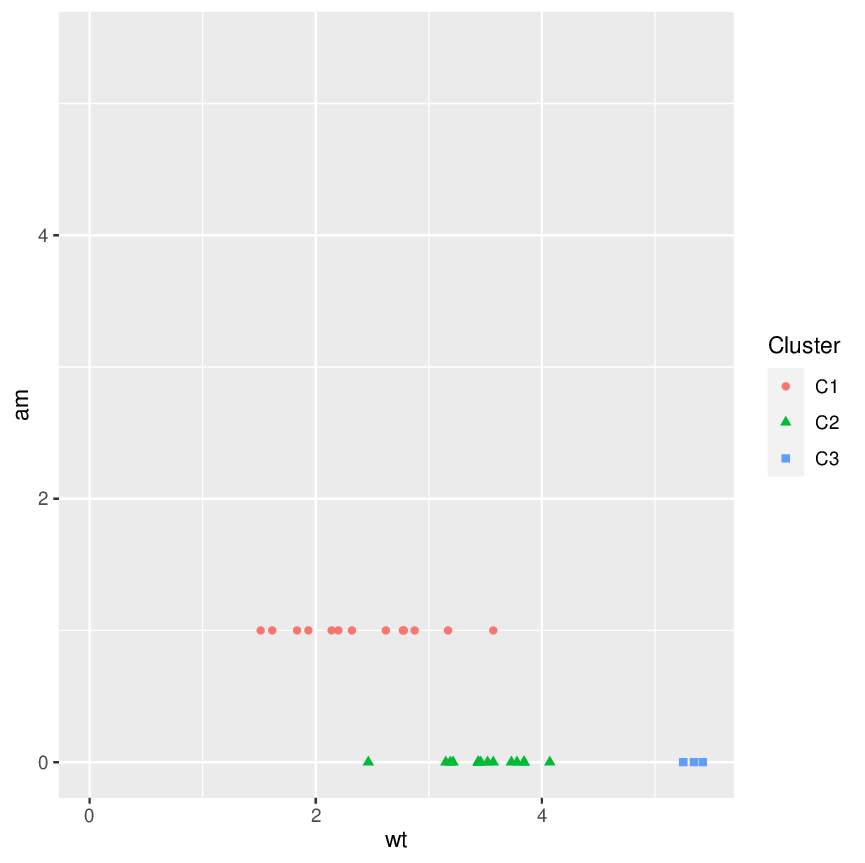}
		\caption{Three clusters C1-C3 resulted in $K$-medoids algorithm for ``mtcars" data set from case-study (6).}\label{f_mtcars}
	\end{figure}

	Clustering results by different indices from Table~\ref{t1:mtcars} shows that the applied $K$-medoids algorithm uncovers three clusters only through our measure. The clusters C1-C3 (Fig.~\ref{f_mtcars}) correspond to three clearly separated groups (see, Table~\ref{t2:mtcars}), where C1 represents the manual transmission cars which have low weights compared to the group of the automatic cars, in which C2 includes the cars with high weights and C3 with the higher ones.
	
	On the other hand, the challengers Dunn and $M_{clus}$ welcome the cluster output from $2$-medoids algorithm, which possesses the first cluster C1 that is the very same as resulted in the $3$-medoids case, and the second cluster by combining C2 with C3 of $3$-medoids method. Therefore, these indices become unable to distinguish the highest-weighted automatic cars from the lower weighted. 
	
	Here we have no information regarding the existing clusters prior to the cluster analysis, whereas the carried out clustering method results in some clusters with much smaller size (e.g. the second cluster of 19 cars splits into clusters of 16 and 3, when we divide the 2-cluster partition into 3) than our previously chosen hyperparameters $J=10$ and $\tilde{K}=9$ in the respective contenders Conn and $\tilde{K}$NN. Therefore, we put $J=2$ and $\tilde{K}=1$ for the present case, where both the contenders come out to be indecisive. 
	
	The rest of the competitors CH and Gap fail with a strictly increasing trend of values as we increase $K$ over the considered range. No doubt, this is a challenging cluster analysis. It has taken its toll on our measure, too. As a result, Table~\ref{t1:mtcars} is showing very close, practically indistinguishable, values of our measure at $K=2$ and at $K=3$, when $l$ is equal to seven; however, a higher value of our index with $l=10$ clearly leads to $\hat{K}=3$, whose robustness is supported by the value corresponding to $l=13$, that increases its reliability with $l=15$.  
	
	Case study (7): Another real-life data set is studied in a trivariate space named ``trees" \footnote{built-in data set `trees' in `R'}, which provides observations on the variables as follows: (i)  girth:	tree diameter in inches, (ii) height: height of tree in ft, and (iii) volume: volume of timber in cubic ft, for 31 felled black cherry trees (Ryan, Joiner and Ryan 1976; Atkinson 1985).
	
\begin{figure}
		\centering
		\includegraphics[width=1\textwidth]{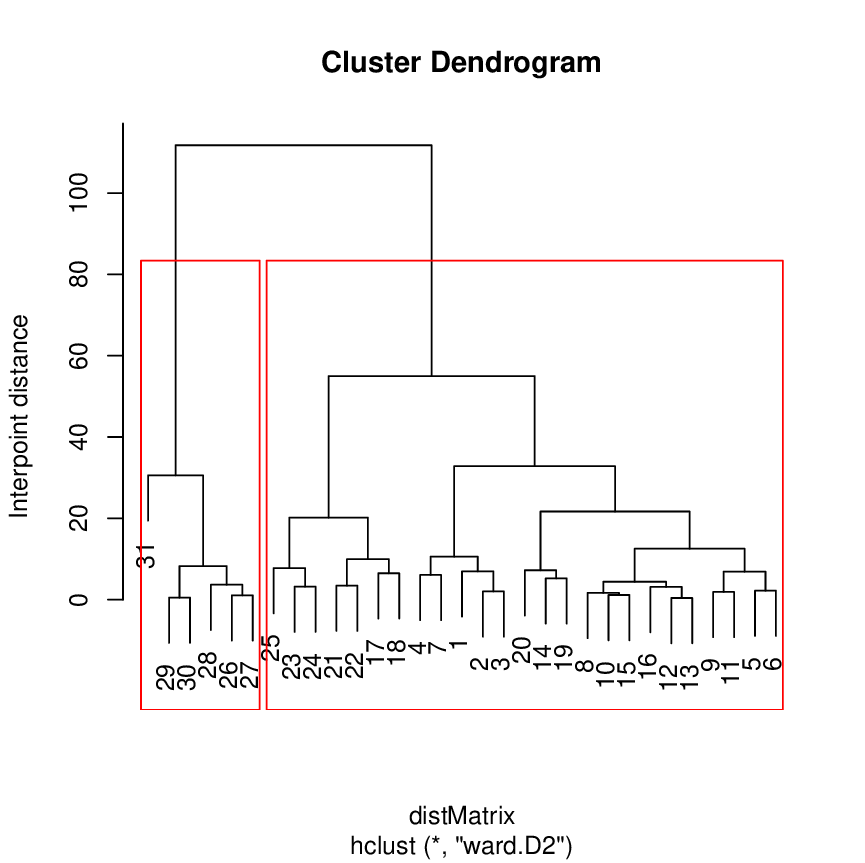}
		\caption{Dendrogram (for singleton clusters, the serial numbers of the members are written at the bottom) resulted in hierarchical clustering with Ward's criterion for ``trees" data set from case-study (7), where two clusters are outlined by red-colored rectangles.}\label{f1_Trees}
	\end{figure}

	Here we implement the agglomerative hierarchical cluster analysis using Ward's criterion (Murtagh and Legendre 2014), which results in different number of clusters indicated by various indices (Table~\ref{t1:trees}). Like the previous case study, to reflect the clustering quality of the smaller resultant clusters accurately in decision making, we fix both the tuning parameters for the competitors Conn and $\tilde{K}$NN at 3. While our index, consistently for all $l$, induces $\hat{K}=2$ supported by Conn and $M_{clus}$ (see, Fig.~\ref{f1_Trees}, for the corresponding dendrogram), Dunn says $\hat{K}=3$ and $\tilde{K}$NN confuses between two and three, whereas CH and Gap give rise to six clusters. 
	
\begin{figure}
		\centering
		\includegraphics[width=1\textwidth]{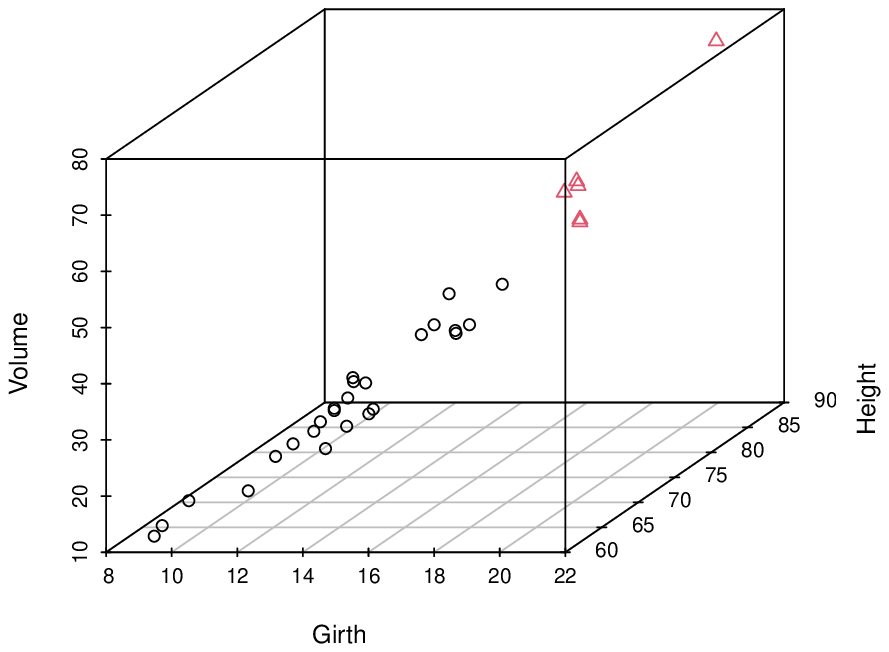}
		\caption{Two clusters for ``trees" data set from case-study (7), obtained through hierarchical clustering, both using Ward's criterion and the average linkage.}\label{f2_Trees}
	\end{figure}

	Such an outcome is not rare in real-world data analyses. Let us solve it by exploring the clusters through detailed investigation (see, Table~\ref{t2:trees}). From the table exhibiting the cluster sizes, the following decision is reached: we discard the choice of $K=6$ for which the resultant partition is actually affected by the outliers present in data due to the deficiency in the hierarchical clustering methods, whereas the cluster properties with respect to each of the study variables are inspected in Table~\ref{t3:trees} for $K=2$ and $3$. It is comprehensible that the first cluster from the former is divided into two further sub-clusters in the latter case. Hence to resolve the conflict around the true number of clusters in this data set, we apply the hierarchical clustering using another criterion, i.e., the average linkage which is generally considered to be much more robust (Kaufman and Rousseeuw 2005). It generates exactly the same results as Ward's characteristic leading to the robustness of quality check through our cluster validity index, and we confirm the verdict of two clusters differentiating trees with lower values for the study variables from the higher ones (see, Fig.~\ref{f2_Trees}). 
	
	Summary: The clustering results of our measure are summarized with respect to the data study performed above. We consider some hypothetical situations in the first five cases to cluster synthetic data, where the proposed index does quiet well compared to its well-known challengers, either outperforming most of them or standing strongly competitive with the best one(s). It is evidenced to be preferred in the single-cluster cases over its only eligible competitor Gap.
	
	Real-life cluster analysis based on the single sample of observations is much more complex to be explored. Therefore, the literature has experienced failure of widely used cluster indices, which are established to be performing good for simulated data sets, while being applied to the challenging data analysis of the real-world samples. Hence, the cluster analysts always suggest the cross-check of the partition, given by the optimal values of such indices, in terms of the detailed exploration of the cluster properties. In what situation a particular measure would supersede the others is unknown, because nothing is known regarding the true clusters underlying the given sample. In this context, we see that our measure executes outstandingly the exposition of the natural groups of data in comparison to its contenders. For example, in the sixth case study, it is proved to be the sole rescuer of the reality, while for the last (i.e. the seventh case study) it consistently, with all considered values of $l$, resolves the conflict arising in the cluster analysis. 
	\section{Conclusion}
	We propose a novel distribution-free clustering accuracy measure to determine the quality of a sample partition. It is based on the multinomial distribution applied to the distances of all members computed from their cluster representatives. With easy construction, simple interpretation, fast computation, good performance, and diverse applicability, it not only estimates the unknown number of clusters, but also checks for the existence of any clustering structure in the designated sample. The latter quality is extremely scarce among the existing cluster validity measures and does not possessed by its considered competitors except the gap statistic. Readers are encouraged to explore the performance of our proposed measure with the selection of different weight functions, and for clustering of high-dimensional data sets with dimension close to or much higher than the sample size. 
	\clearpage 
	\section*{Appendix: \\R codes to compute our measure $C^K_{MN}$} 
	\verb!##! Codes are illustrated below for clustering the first synthetic data set from Case-study (1), through $K$-medoids algorithm, whose output is printed in Table~\ref{t1:MulNor} (i.e. second column in the matrix of results under $K$-medoids)\\\\
	\verb!##! Loading library `cluster' for implementing $K$-medoids clustering through `PAM' algorithm\\
	\verb!library(cluster)!\\
	\verb!##! Loading library `copula' for generating data under multivariate set-up constructed by a copula\\
	\verb!library(copula)!\\\\ 
	\verb!##!--------Generation of data set--------\\
	\verb!##! Dimension of the data set\\
	\verb!p<-10!\\
	\verb!##! Specify copula\\
	\verb!myCop.t<-ellipCopula("t",param=.15,dim=p,dispstr="ex",df=2)!\\ 
	\verb!##! Define population distribution\\
	\verb!myMvd<-mvdc(copula=myCop.t,margins=rep("norm",p),!\\ 
	\verb!paramMargins=sapply(1:p,function(j)list(list(mean=0,sd=1))))!\\
	\verb!##! Assign cluster sizes and thereby compute total sample size\\
	\verb!N1<-45;N2<-50;N3<-70;N<-N1+N2+N3!\\ 
	\verb!##! Fix data for reproducible output\\
	\verb!set.seed(12045)!\\  
	\verb!##! Draw sample of size N from the population\\ 
	\verb!Z<-rMvdc(N,myMvd)!\\
	\verb!data<-matrix(NA,N,p)!\\
	\verb!##! Sample from first cluster \\
	\verb!data[1:N1,]<-Z[1:N1,]!\\ 
	\verb!##! Sample from second cluster \\
	\verb!data[(N1+1):(N1+N2),]<-Z[(N1+1):(N1+N2),]!\\
	\verb!+matrix(rep(-3,N2),N2,p,byrow=T)!\\
	\verb!##! Sample from third cluster \\
	\verb!data[(N1+N2+1):(N1+N2+N3),]<-Z[(N1+N2+1):(N1+N2+N3),]!\\
	\verb!+matrix(rep(3,N3),N3,p,byrow=T)!\\
	\verb!##! Compute interpoint distances for every unique pair of observations from entire sample, by default it is Euclidean metric \\
	\verb!distMatrix<-dist(data)!\\
	\verb!##! Create $N\times N$ distance matrix \\
	\verb!M<-as.matrix(distMatrix)!\\\\
	\verb!##!------Hyperparameters specification--------\\
	\verb!##! Max no. of clusters considered\\
	\verb!K_max<-6!\\  
	\verb!##! Value of hyperparameter $l$ in our measure\\
	\verb!l<-10!\\ 
	\verb!##! Sub-intervals for distances to computes counts: N\_j$|$k s \\
	\verb!interval<-seq(0,1,1/l)!\\\\ 
	\verb!##!------Construction of our measure--------\\
	\verb!measure<-NULL!\\
	\verb!##! K loop begins\\
	\verb!for(K in 2:K_max){!\\ 
		\verb!stat<-NULL!\\
		\verb!##! Cluster data by PAM algorithm\\
		\verb!kmed<-pam(distMatrix,k=K,diss=T)!\\
		\verb!##! Cluster memberships\\ 
		\verb!cl<-kmed$clustering!\\
		\verb!##! Cluster representatives' id\\   
		\verb!Rk_id<-kmed$id.med!\\\\  
		\verb!N_j<-matrix(NA,K,l)!\\
		\verb!##! k loop begins\\
		\verb!for(k in 1:K){!\\ 
			\verb!Sk<-NULL!\\
			\verb!##! Compute the set from Eq.~(1)\\
			\verb!Sk<-M[which(cl==k),Rk_id[k]]!\\
			\verb!##! Compute the set from Eq.~(2)\\
			\verb!Sk_norm<-Sk/max(Sk)!\\
			\verb!##! Compute variables from Eq.~(3)\\
			\verb!N_j[k,]<-hist(Sk_norm,breaks =interval,plot=F)$counts!\\
			\verb!##! k loop ends\\
			\verb!}!\\\\
		\verb!##! Compute vector of variables from Eq.~(4)\\
		\verb!Nk<-apply(N_j,1,sum)!\\
		\verb!##! Give estimated variance terms from Eq.~(12)\\
		\verb!term<-function(x,k){return((x/Nk[k])* (Nk[k]-x))}!\\
		\verb!##! k loop begins again\\
		\verb!for(k in 1:K){!\\  
			\verb!objective.in <- term(N_j[k,],k)!\\
			\verb!wt<-rep(NA,l)!\\ 
			\verb!for(j in 1:l)!\\
			\verb!##! Assign decreasing weights\\
			\verb!wt[j]<-(l+1-j)!\\
			\verb!##! Compute cluster-wise terms in Eq.~(13)\\   
			\verb!stat[k]<-wt%*%objective.in!\\  
			\verb!##! k loop ends again\\
			\verb!}!\\ 
		\verb!##! Calculate our measure as per Eq.~(13)\\
		\verb!measure[K]<-sum(stat)!\\ 
		\verb!##! K loop ends\\
		\verb!}!\\\\ 
	\verb!##!--------Output--------\\	
	\verb!K_hat<-which.max(measure[-1])+1 !\\
	\verb!##! Print estimated number of clusters \\
	\verb!K_hat!\\  
	\verb!##! Report computed values of our measure for different number of clusters\\
	\verb!round(measure,5)!\\

	\clearpage   
	\begin{table}
		\caption{Computed values of different indices for a multivariate normal data set from case-study (1), through $K$-means and $K$-medoids algorithms, with varying $K$ (Original $K$ and optimal values of indices are marked in bold).} 
		\begin{center}
			\begin{tabular}{cccccccc}
				\hline\\
				$K$&$C^K_{MN}$&Dunn&Conn&CH&$\tilde{K}$NN&Gap&$M_{clus}$\\
				&$(l=10)$&&$(J=10)$&&($\tilde{K}=9$)&&($h=h^{*}$)\\
				\hline\\	
				&&&&$K$-means&&&\\	\hline\\
				2& 741.7756& 0.12315& 12.57500& 349.4184&  1.81818& 0.4412529& 0.5418888\\[1ex]
				\textbf{3}& \textbf{828.0040}& \textbf{0.31933}&  \textbf{2.06627}& \textbf{450.0298}&  \textbf{0}& \textbf{0.6611909}& \textbf{0.5862262}\\[1ex]
				4& 801.7292& 0.15702& 31.05952& 329.0211&  3.63636& 0.6531145& 0.4549353\\[1ex]
				5& 778.9145& 0.13795& 62.54841& 258.5651&  9.69697& 0.6480189& 0.3020418\\[1ex]
				6& 767.1396& 0.13795& 85.91746& 210.8898& 12.72727& 0.6351098& 0.2841469\\[1ex]
				
				\hline\\	
				&&&&$K$-medoids&&&\\	\hline\\
				2&738.1775& 0.09814& 32.10437& 333.6501&  0.60606& 0.4206209& \textbf{0.6204327}\\[1ex]
				\textbf{3}& \textbf{820.8229}& \textbf{0.31229}&  \textbf{4.33333}& \textbf{446.6430}& \textbf{ 1.21212}& \textbf{0.6717865}& 0.5854474\\[1ex]
				4& 801.6123& 0.10607& 50.06786& 313.2806&  4.24242& 0.6541694& 0.4420826\\[1ex]
				5&791.6566& 0.10907& 58.15992& 252.4601&  4.84848& 0.6596946& 0.3689428\\[1ex]
				6&784.2477& 0.10907& 90.00040& 211.4078& 12.12121& 0.6591167& 0.2241668\\[1ex]
				
				\hline
			\end{tabular}
		\end{center}
		\label{t1:MulNor}
	\end{table}  
	\clearpage   
	\begin{table}
		\caption{Original misclassification rates for clustering results of a multivariate normal data set from case-study (1), having $K=3$ clusters, through $K$-means and $K$-medoids algorithms (Results from better algorithm are highlighted in bold).} 
		\begin{center}
			\begin{tabular}{ccc}
				\hline\\
				Algorithm&$C^3_{MN}$&Misclassification\\
				&   $(l=10)$       & rate (\%)\\
				\hline\\
				$K$-means&\textbf{828.0040}&\textbf{0.606}\\[1ex]
				$K$-medoids&820.8229&1.818\\[1ex]
				\hline
			\end{tabular}
		\end{center}
		\label{t2:MulNor}
	\end{table}  
	\clearpage   
	\begin{table}
		\caption{Clustering results for a multivariate normal data set from case-study (1), in terms of  $C^K_{MN}$, for different $l$, through $K$-means algorithm with varying $K$ (Original $K$ and optimal values of $C^K_{MN}$ are written in bold).} 
		\begin{center}
			\begin{tabular}{cccccc}
				\hline\\
				$K$&2&\textbf{3}&4&5&6\\[1ex]
				
				\hline\\
				$C^K_{MN}(l=5)$&310.9756& \textbf{350.0653}& 330.2694& 330.0142& 321.2280\\[1ex]
				
				\hline\\
				$C^K_{MN}(l=7)$&485.3022& \textbf{554.9200}& 537.1359& 517.5058& 512.6104\\[1ex]
				
				\hline\\
				$C^K_{MN} (l=10)$&741.7756& \textbf{828.0040}& 801.7292& 778.9145& 767.1396\\[1ex]
				
				\hline
			\end{tabular}
		\end{center}
		\label{t3:MulNor}
	\end{table} 
	\clearpage   
	\begin{table}
		\caption{Clustering results by various indices, in terms of the number of correctly identifying $K$ out of 1000  Monte Carlo replications, under different case studies.} 
		\begin{center}
			\begin{tabular}{c|ccccccc}
				\hline\\
				Case &$C^K_{MN}$ $(l)$ &Dunn&Conn&CH&$\tilde{K}$NN&Gap&$M_{clus}$\\
				study	&&&$(J=10)$&&($\tilde{K}=9$)&&($h=h^{*}$)\\[1ex]
				\hline\\
				(1)&950 $(13)$&942&938&999& 834&981&854\\[1ex]
				\hline
				(2)&968 (13)&--&--&--&--&968&--\\[1ex]
				\hline
				(3)&703 $(30)$&116&0&798&0&636&758\\[1ex]
				\hline
				(4)&974 (30)&--&--&--&--&724&--\\[1ex]
				\hline
			\end{tabular}
		\end{center}
		\label{t:MCS}
	\end{table}   
	\clearpage    
	\begin{table}
		\caption{Clustering results through $K$-means algorithm for a single-cluster multivariate normal data set from case-study (2), with varying $K$ (Original $K$ and optimal value of $C^K_{MN}$ are written in bold).} 
		\begin{center}
			\begin{tabular}{cc}
				\hline\\
				$K$&$C^K_{MN}(l=10)$\\
				\hline\\
				\textbf{1}& \textbf{874.5454}\\[1ex]
				2&821.6158 	\\[1ex]
				3&753.2995 	\\[1ex]
				4&759.0228 	\\[1ex]
				5&734.6703 	\\[1ex]
				6&698.1634	\\[1ex]
				\hline
			\end{tabular}
		\end{center}
		\label{t:noclustering}
	\end{table}  
	\clearpage
	\begin{sidewaystable}
		\centering
		\caption{Clustering results through DBSCAN for arbitrary-shaped noisy clusters from case-study (5) (Better outcomes are highlighted in bold).}
		\begin{tabular}{cccccccccc}
			\hline\\
			Parameters&$\hat{K}$&$C^{\hat{K}}_{MN}$&$C^{\hat{K}}_{MN}$&Dunn&Conn&CH&$\tilde{K}$NN&Gap&$M_{clus}$\\
			of DBSCAN&&$(l=10)$&$(l=10)$&&$(J=10)$&&$(\tilde{K}=9)$&&($h=h^{*}$)\\
			&&(mean)&(mode)&&&\\
			\hline\\
			$Minpts=4,\epsilon=0.15$&5&2826.422&2934.361&0.04594&17.83016&24.80163&1.02916&--& -0.20706\\[1ex]
			$Minpts=4,\epsilon=0.17$&\textbf{4}& \textbf{2907.455}&\textbf{2966.377}&\textbf{0.04762}
			&\textbf{13.12143}&\textbf{32.92533}&\textbf{0.84175}&--& \textbf{-0.18839}
			\\[1ex]
			\hline
		\end{tabular}
		\label{t:ArShCl} 
	\end{sidewaystable}
	\clearpage
	\begin{sidewaystable}
		\centering
		\caption{Computed values of different indices, for ``mtcars" data set from case-study (6), through $K$-medoids clustering with varying $K$ (Optimal values of indices are marked in bold).} 
		\begin{tabular}{ccccccccccc}
			\hline\\
			$K$&$C^K_{MN}$&$C^K_{MN}$&$C^K_{MN}$&$C^K_{MN}$&Dunn&Conn&CH&$\tilde{K}$NN&Gap&$M_{clus}$\\
			&$(l=7)$&$(l=10)$&$(l=13)$&$(l=15)$&&($J=2$)&&($\tilde{K}$=1)&&($h=h^{*}$)\\
			\hline\\	
			2&\textbf{114.93117}&168.6883&229.0850&262.6073&\textbf{1.32173}&\textbf{0.0}& 420.7910&\textbf{0.000}& 0.14600&\textbf{0.88644}
			\\[1ex]
			3&114.28205&\textbf{170.2003}&\textbf{230.5401}&\textbf{269.5897}& 0.57365&\textbf{0.0}& 491.0951&\textbf{0.000}& 0.33125& 0.85514\\[1ex]
			4&105.61905&161.6875&218.9375&250.9762& 0.18692&\textbf{0.0}& 657.8667&\textbf{0.000}& 0.50062&0.75078
			\\[1ex]
			5&96.90996&147.8182&202.8636&232.1398& 0.14480& \textbf{0.0}&704.0130&\textbf{0.000}&0.59624&0.65114\\[1ex]
			6&93.71905&137.0000&174.6000&211.5762& 0.16842& 1.5&\textbf{836.9777}&3.125&\textbf{0.65425}&0.68104\\[1ex]
			\hline
		\end{tabular}
		\label{t1:mtcars}
	\end{sidewaystable}
	\clearpage
	\begin{table}
		\caption{Clustering results, for ``mtcars" data set from case-study (6), through $K$-medoids algorithm with $K=3$, where the value on wt and the category on am are given for each member from individual clusters C1-C3.} 
		\begin{center}
			\begin{tabular}{cc|cc|cc}
				\hline
				Cluster&C1&Cluster&C2&Cluster&C3\\
				\hline
				wt  &    am &wt&am&wt&am\\
				\hline
				
				2.620& manual& 3.215& automatic&5.250& automatic\\[1ex]
				2.875& manual&3.440& automatic&5.424& automatic\\[1ex]
				2.320& manual&3.460& automatic&5.345& automatic\\[1ex]
				2.200& manual&3.570& automatic&&\\[1ex]
				1.615& manual&3.190& automatic&&\\[1ex]
				1.835& manual&3.150& automatic&&\\[1ex]
				1.935& manual&3.440& automatic&&\\[1ex]
				2.140& manual&3.440& automatic&&\\[1ex]
				1.513& manual&4.070& automatic&&\\[1ex]
				3.170& manual&3.730& automatic&&\\[1ex]
				2.770& manual&3.780& automatic&&\\[1ex]
				3.570& manual&2.465& automatic&&\\[1ex]
				2.780& manual&3.520& automatic&&\\[1ex]
				&       &3.435& automatic&&\\[1ex]
				&       &3.840& automatic&&\\[1ex]
				&       &3.845& automatic&&\\[1ex]
				\hline
			\end{tabular}
		\end{center}
		\label{t2:mtcars}
	\end{table} 
	\clearpage
	\begin{sidewaystable}
		\centering
		\caption{Computed values of different indices, for ``trees" data set from case-study (7), through hierarchical clustering (Ward's criterion) with varying $K$ (Optimal values of indices are marked in bold).} 
		\begin{tabular}{ccccccccccc}
			\hline\\
			$K$&$C^K_{MN}$&$C^K_{MN}$&$C^K_{MN}$&$C^K_{MN}$&Dunn&Conn&CH&$\tilde{K}$NN&Gap&$M_{clus}$\\
			&$(l=5)$&$(l=7)$&$(l=10)$&$(l=13)$&&($J=3$)&&($\tilde{K}=3$)&&($h=h^{*}$)\\
			\hline\\	
			2&\textbf{74.94667}&\textbf{107.3533}& \textbf{155.62}&\textbf{193.4733}& 0.25164&\textbf{0.00000}& 53.71202&\textbf{0.00000}& 0.14632& \textbf{0.63335}\\[1ex]
			3&72.07143&92.16667&139.8175&168.1429	&\textbf{0.29398}&0.33333& 58.36777&\textbf{0.00000}&0.16495& 0.54648\\[1ex]
			4&69.19194&102.418&144.1747&177.5146&0.13065& 3.16667& 56.47167&3.22581& 0.18346& 0.51240 \\[1ex]
			5&63.92527&90.58462&129.6747&165.8813&0.17736& 5.00000& 66.63073& 6.45161 &0.19946&0.49513\\[1ex]
			6&50.17143&68.20000&98.12857&126.5762&0.19692&7.33333& \textbf{72.57006}&12.90323&\textbf{0.22338}& 0.46627
			\\[1ex]
			\hline
		\end{tabular}
		\label{t1:trees}
	\end{sidewaystable}   
	\clearpage
	\begin{table}
		\caption{Clustering results for ``trees" data set form case-study (7) through hierarchical clustering (Ward's criterion) with varying $K$.} 
		\begin{center}
			\begin{tabular}{cc}
				\hline\\
				$K$&Cluster\\
				&sizes\\
				\hline\\	
				2& $(25,6)$\\[1ex]
				3&$(18,7, 6)$\\[1ex]
				4&$(5,13,7,6)$\\[1ex]
				5& $(5,13,7,5,1)$\\[1ex]
				6&$(5, 10,3,7,5,1)$\\[1ex]
				\hline
			\end{tabular}
		\end{center}
		\label{t2:trees}
	\end{table}   
	\clearpage
	\begin{table}
		\caption{Cluster properties for ``trees" data set from case-study (7) through hierarchical clustering (Ward's criterion) with $K=2$ and $3$.} 
		\begin{center}
			\begin{tabular}{ccccc}
				\hline\\
				$K$&Cluster&&Cluster means&\\
				&no.&&(standard errors)&\\
				&&Girth &  Height &  Volume\\
				\hline\\	
				2& 1& 12.056& 74.640& 23.456 \\[1ex] 
				&          &(2.045107)& (6.137622)& (8.526668) \\[1ex]
				&2&18.21667& 81.66667& 58.15000\\[1ex]
				&         &(1.097598)& (2.494438)& (8.796353)\\[1ex]
				\hline
				3&1&11.12222& 73.00000 &18.98889 \\[1ex]
				&&(1.466119)& (5.792716)& (4.679256) \\[1ex]
				&2&14.45714& 78.85714& 34.94286 \\[1ex]
				&&	(1.184250)& (4.852939)& (4.482619) \\[1ex]
				&3&18.21667& 81.66667& 58.15000 \\[1ex]
				&&	(1.097598)& (2.494438)& (8.796353)\\[1ex]
				\hline
			\end{tabular}
		\end{center}
		\label{t3:trees}
	\end{table}   
	\clearpage   
	\clearpage
	\clearpage
	\clearpage
	\clearpage  
	\clearpage   
	{}

\clearpage
	\begin{thebibliography}{}
		\bibitem{}
		Agresti, A. (2002). \textsl{Categorical Data Analysis}. John Wiley \& Sons, Inc., New Jersey.
		\bibitem{}
		Aljarah, I. and Ludwig, S. A. (2013) \textsl{A new clustering approach based on glowworm swarm optimization}. IEEE congress on evolutionary computation.
		\bibitem{}
		Atkinson, A. C. (1985). \textsl{Plots, Transformations and Regression}. Oxford University Press.
		\bibitem{}
		Balcan, M.-F., Liang, Y. and Gupta, P. (2014). \textsl{Robust Hierarchical Clustering}. Journal of Machine Learning Research, \textbf{15} 4011--4051.
		\bibitem{}	
		Bandyopadhyay, U. \& Modak, S. (2018). \textsl{Bivariate density estimation
			using normal-gamma kernel with application to astronomy}. Journal of Applied Probability and Statistics. \textbf{13}, 23--39.
		\bibitem{}
		Cali\'{n}ski, T. \&  Harabasz, J. (1974). \textsl{A Dendrite Method for Cluster Analysis}. Communications in Statistics -- Theory and Methods. \textbf{3}, 1--27.
		\bibitem{}
		Dobson, A. J. (1990). \textsl{An Introduction to Generalized Linear Models}. Chapman and Hall, London.
		\bibitem{}
		Dunn, J. C. (1974). \textsl{Well-separated clusters and optimal fuzzy partitions.}  Journal of Cybernetics. \textbf{4}, 95--104.
		\bibitem{}
		Ester, M., Kriegel, H.-P., Sander, J. \& Xu, X. (1996). \textsl{A density-based algorithm for discovering clusters
			in large spatial databases with noise.} Proceedings of the Second International Conference on
		Knowledge Discovery and Data Mining (KDD-96). AAAI Press, Portland, Oregon, 226--231.
		\bibitem{}
		Everitt, B. S., Landau, S. and Leese,  M. (2001). \textsl{Cluster Analysis.} Arnold, London.
		\bibitem{}
		Frayley, C. and Raftery, A. E. (1998), \textsl{How Many Clusters? Which Clustering
			Method? Answers via Model-Based Cluster Analysis}. The Computer Journal. \textbf{41}, 578--588.
		\bibitem{}
		Handl, J., Knowles, K. \& Kell, D. (2005). \textsl{Computational cluster validation in post-genomic data analysis}. Bioinformatics. \textbf{21}, 3201--3212.
		\bibitem{}
		Hahsler, M., M. Piekenbrock, and D. Doran. (2019). \textsl{dbscan: Fast density-based clustering with R.} Journal of
		Statistical Software. \textbf{9}, 1--30.
		\bibitem{}
		Hartigan, J. A. \&  Wong, M. A. (1979). \textsl{A K-means clustering algorithm.} Applied Statistics. \textbf{28}, 100--108. 
		\bibitem{}
		Jain, A. K., Murty, M. N. and Flynn, P. J. (1999).  \textsl{Data clustering: a
			review.} ACM Computing Surveys. 31, 264–323.
		\bibitem{}
		Joanes, D. N. and Gill, C. A. (1998). \textsl{Comparing measures of sample skewness and kurtosis}. The Statistician, \textbf{47}, 183--189.
		\bibitem{}
		Johnson, N. L., Kotz, S. and Balakrishnan, N. (1997). \textsl{Discrete Multivariate Distributions}. John Wiley \& Sons, Inc.,
		New York.
		\bibitem{}
		Johnson, R. A. and Wichern, D. W. (2007). \textsl{Applied Multivariate 
			Statistical Analysis}, Pearson Prentice Hall, New Jersey.
		\bibitem{}
		Kaufman, L. and Rousseeuw, P. J. (2005). \textsl{Finding Groups in Data: An Introduction to Cluster Analysis.} John Wiley and Sons, New Jersey.
		\bibitem{}
		Lewis, D. D. and Gale W. A. (1994). \textsl{A sequential algorithm for training text classifiers.} In: Proceedings of the 17th annual international ACM SIGIR conference on research
		and development in information retrieval.
		\bibitem{}
		MacQueen, J. (1967). \textsl{Some methods for classification and analysis of multivariate observations}. In Proceedings of the Fifth Berkeley Symposium on Mathematical Statistics and Probability, eds L. M. Le Cam \& J. Neyman, 1, pp. 281–297. Berkeley, CA: University of California Press.
		\bibitem{}
		Matioli, L. C., Santos,  S. R., Kleina,  M. \& Leite, E. A. (2018). \textsl{A new algorithm for clustering based on kernel density estimation}. Journal of Applied Statistics. \textbf{45}, 347--366.
		\bibitem{}
		McLachlan, G. and Peel, D. (2000). \textsl{Finite Mixture Models}. John Wiley and Sons,
		New York.
		\bibitem{}
		McCullagh, P. and Nelder, J. A. (1989). \textsl{Generalized Linear Models}. Chapman and Hall, London.
		\bibitem{}
		Modak, S. (2019). \textsl{Uncovering astrophysical phenomena related to galaxies and other objects through statistical analysis.} Ph.D. Thesis, University of Calcutta, India. URL: http://hdl.handle.net/10603/314773 
		\bibitem{}
		Modak, S. (2021). \textsl{Distinction of groups of gamma-ray bursts in the BATSE catalog through fuzzy clustering}. Astronomy and Computing. \textbf{34}, Article id 100441, 1--7.
		\bibitem{}
		Modak, S. (2022). \textsl{A new nonparametric interpoint distance-based measure for assessment of clustering}. Journal of Statistical Computation and Simulation. \textbf{9},	1062--1077.
		\bibitem{}
		Modak, S. (2023a). \textsl{A new interpoint distance-based clustering algorithm using kernel density estimation}. Communications in Statistics -- Simulation and Computation. In Press, Doi: 10.1080/03610918.2023.2179071
		\bibitem{}
		Modak, S. (2023b), \textsl{Pointwise norm-based clustering of data in arbitrary dimensional space}, Communications in Statistics - Case Studies, Data Analysis and Applications. \textbf{9}, 121–134.
		\bibitem{}
		Modak, S. (2023c), \textsl{Validity index for clustered data in non-negative space}, Calcutta Statistical Association Bulletin, \textbf{75}, 60–71.
		\bibitem{}
		Modak, S. (2023d). \textsl{A Book Review on ``Finding Groups in Data: An Introduction to Cluster Analysis by Kaufman, L. and Rousseeuw, P. J. (2005)".} Journal of Applied Statistics, DOI: https://doi.org/10.1080/02664763.2023.2220087
		\bibitem{}
		Modak, S. (2023e). \textsl{A new measure for assessment of clustering based on kernel	density estimation}. Communications in Statistics -- Theory and Methods. 52, 5942-5951.	
		\bibitem{}
		Modak, S. (2023f). \textsl{A Book Review of Astrostatistical Fundamentals:
			Statistical Methods for Astronomical Data Analysis authored by
			Asis Kumar Chattopadhyay and Tanuka Chattopadhyay, 2014}. DOI: https://doi.org/10.1080/02664763.2023.2220087.	
		\bibitem{}
		Modak, S. (2023g). \textsl{Determination of the number of clusters through logistic regression analysis}. Journal of Applied Statistics, DOI: https://doi.org/10.1080/02664763.2023.2283687.
		\bibitem{}
		Modak, S. \& Bandyopadhyay, U. (2019). \textsl{A new nonparametric test for
			two sample multivariate location problem with application to astronomy}. Journal of Statistical Theory and Applications. \textbf{18}, 136--146.
		\bibitem{}
		Modak, S., Chattopadhyay, A. K. \& Chattopadhyay, T. (2018). \textsl{Clustering of gamma-ray bursts through kernel principal component analysis}. Communications in Statistics -- Simulation and Computation. \textbf{47}, 1088--1102.
		\bibitem{}
		Modak, S., Chattopadhyay, T. \& Chattopadhyay, A. K. (2017). \textsl{Two
			phase formation of massive elliptical galaxies: study through cross--correlation including spatial effect.} Astrophysics and Space Science. \textbf{362}, Article id: 206, pages 1--10.
		\bibitem{}
		Modak, S., Chattopadhyay, T. \& Chattopadhyay, A. K. (2020). \textsl{Unsupervised classification of eclipsing binary light curves through k-medoids
			clustering}. Journal of Applied Statistics. \textbf{47}, 376--392.
		\bibitem{}
		Modak, S., Chattopadhyay, T. \& Chattopadhyay, A. K. (2022). \textsl{Clustering of eclipsing binary light curves through functional principal component analysis}. Astrophysics and Space Science. \textbf{ 367}, Article id: 19, pages 1--10
		\bibitem{}
		Murtagh, Fionn and Legendre, Pierre (2014). \textsl{Ward's hierarchical agglomerative clustering method: which algorithms implement Ward's criterion?} Journal of Classification, \textbf{31}, 274–295.
		\bibitem{}
		Pakhira, M. K., Bandyopadhyay, S. and Maulik, U. (2004). \textsl{
			Validity index for crisp and fuzzy clusters}. Pattern Recognition, \textbf{37}, 487-501.
		\bibitem{}
		Ryan, T. A., Joiner, B. L. and Ryan, B. F. (1976). \textsl{The Minitab Student Handbook}. Duxbury Press.
		\bibitem{}
		Rousseeuw, P. J. (1987). \textsl{Silhouettes: A graphical aid to the interpretation and validation of cluster analysis.} Journal of Computational and Applied Mathematics. \textbf{20}, 53--65. 
		\bibitem{}
		Ripley, B. D. (1996). Pattern Recognition and Neural Networks. Cambridge University Press, Cambridge.
		\bibitem{}
		Rousseeuw, P. J. (1987). \textsl{Silhouettes: A graphical aid to the interpretation
			and validation of cluster analysis}. Journal of Computational and Applied
		Mathematics. \textbf{20}, 53--65.
		\bibitem{}	
		Ruspini, E. H.  (1970). \textsl{Numerical methods for fuzzy clustering.} Information Sciences. \textbf{2}, 319--350.
		\bibitem{}
		Schwarz, G. (1978). \textsl{Estimating the Dimension of a Model.} The Annals of Statistics, \textbf{6}, 461--464.
		\bibitem{}
		Silva, L. E. B. D., Melton, N. M. and  Wunsch, D. C. (2020). \textsl{Incremental Cluster Validity Indices for Online Learning of Hard Partitions: Extensions and Comparative Study}. in IEEE Access, \textbf{8}, 22025-22047.
		\bibitem{}
		Silverman, B. W. (1986), {\it Density Estimation for Statistics and Data Analysis}, Chapman
		and Hall, London.
		\bibitem{}
		Sureja, N., Chawda, B. and Vasant, A. (2022). \textsl{
			An improved K-medoids clustering approach based on the crow search algorithm}.
		Journal of Computational Mathematics and Data Science,
		\textbf{3},
		100034.
		\bibitem{}
		Tibshirani, R., Walther, G. and Hastie, T. (2001). \textsl{Estimating the number of data clusters via the Gap statistic}. Journal of the Royal Statistical Society B, \textbf{63}, 411--423.
		\bibitem{}
		Vale, D. C. and Maurelli V. A. (1983). \textsl{Simulating multivariate nonnormal distributions}. Psychometrika, \textbf{48}, 465--471.
	\end{thebibliography}
\end{document}